\documentclass[usenatbib,fleqn]{mn2e}
\usepackage{fixltx2e,longtable,ifpdf,amsmath,amssymb}
\usepackage{times,nicefrac}
\ifpdf
  \usepackage{aeguill}
  \usepackage[pdftex]{graphicx,color}
\else
  \usepackage[dvips]{graphicx,color}
\fi

\newcommand{\kms} {\ensuremath{\mbox{km}\;\mbox{s}^{-1}}}
\newcommand{\pd}  {\ensuremath{\mbox{d}^{-1}}}

\def\fh{\hbox{$.\!\!^{\rm h}$}}
\def\fm{\hbox{$.\!\!^{\rm m}$}}

\def\utw{\smash{\rlap{\lower5pt\hbox{$\sim$}}}}
\def\udtw{\smash{\rlap{\lower6pt\hbox{$\approx$}}}}

\title[Satellite photometry of $\zeta$~Oph]{Amplitude variability in
  satellite photometry of the non-radially pulsating O9.5$\,$V star
  $\zeta$~Oph} \author[I. D. Howarth et al. ]{Ian D. Howarth,$^1$\thanks{i.howarth@ucl.ac.uk}
  K. J. F. Goss$^{2}$, I. R. Stevens$^{2}$, W. J. Chaplin$^{2}$, and
  Y. Elsworth$^{2}$\\ $^{1}$Dept. Physics \& Astronomy, University
  College London, Gower St., London WC1E 6BT \\ $^{2}$School of
  Physics and Astronomy, University of Birmingham, Edgbaston,
  Birmingham B15 2TT}

\begin{document}

\pagerange{\pageref{firstpage}--\pageref{lastpage}} \pubyear{}

\maketitle

\label{firstpage}

\begin{abstract}
  We report a time-series analysis of satellite photometry of the
  non-radially pulsating Oe star $\zeta$~Oph, principally using data
  from \emph{SMEI} obtained 2003--2008, but augmented with \emph{MOST}
  and \emph{WIRE} results. Amplitudes of the strongest photometric
  signals, at 5.18, 2.96, and 2.67~\pd, each vary independently over
  the $5\nicefrac{1}{2}$-year monitoring period (from $\sim$30 to
  $\lesssim$2~mmag at 5.18~\pd), on timescales of hundreds of days.
  Signals at 7.19~\pd\ and 5.18~\pd\ have persisted (or recurred) for 
  around two decades.  Supplementary
  spectro\-scopic observations show an H$\alpha$ emission episode in
  2006; this coincided with small increases in amplitudes of the three
  strongest photometric signals.

\end{abstract}


\section{Introduction}
\label{sec:intro}

$\zeta$~Oph\footnote{HD 149757} is the nearest O-type star, and one of the brightest ($\pi
= 8.91\pm0.20$~mas, \citealt{vanLeeuwen07}; O9.5$\,$Vnn, $V \simeq
2.6$, \citealt{Sota11}).  
Detailed spectro\-scopic investigations of
velocity-resolved absorption-line structure have been facilitated by
its brightness and exceptionally rapid
rotation ($v_{\rm e}\sin{i} \gtrsim 400$~\kms;
\citealt{Howarth01}, \citealt{Villamariz05}); periodic line-profile
variability, discovered by \citet{Walker79}, has subsequently been
widely interpreted in terms of non-radial pulsations (NRP;
\citealt{Vogt83}, \citealt{Reid93}, \citealt{Kambe97}).  H$\alpha$
emission episodes lasting, typically, several weeks have been observed
to occur every few years (e.g., \citealt{Ebbets81, Kambe93}); the
inferred circumstellar decretion disk is probably causally associated
with rapid rotation, and possibly with NRP \citep{Cranmer09}, though
the latter remains a open \mbox{issue}.

Spectro\-scopic line-profile variability associated with NRP is
primarily sensitive to sectoral pulsation modes (since tesseral modes
readily lead to cancellation in velocity space). Moreover, the
requirements of high signal-to-noise ratio and high resolution typically limit
spectro\-scopic time series to only a few nights, resulting in further
detection biases, towards short periods and large amplitudes. These
observational constraints contrast with satellite-based photo\-metry,
which can yield precise measurements over an extended time period,
thereby affording the opportunity to investigate pulsation characteristics
in a parameter space inaccessible to spectro\-scopic study.  For
$\zeta$~Oph, this opportunity was exploited by \citet{Walker05}, who
found a number of periodic signals in 24~days of high-cadence,
near-continuous photo\-metry from the \emph{Microvariability and
  Oscillations of STars (MOST)} satellite.  Here we report new results
from the \emph{Solar Mass Ejection Imager (SMEI)} and \emph{Wide-field
  Infra\-Red Explorer (WIRE)} missions; for completeness, we also
include our independent re-analysis of the \emph{MOST} data.  

\section{Observations}

\begin{figure}
\centering
\includegraphics[scale=0.32,angle=270]{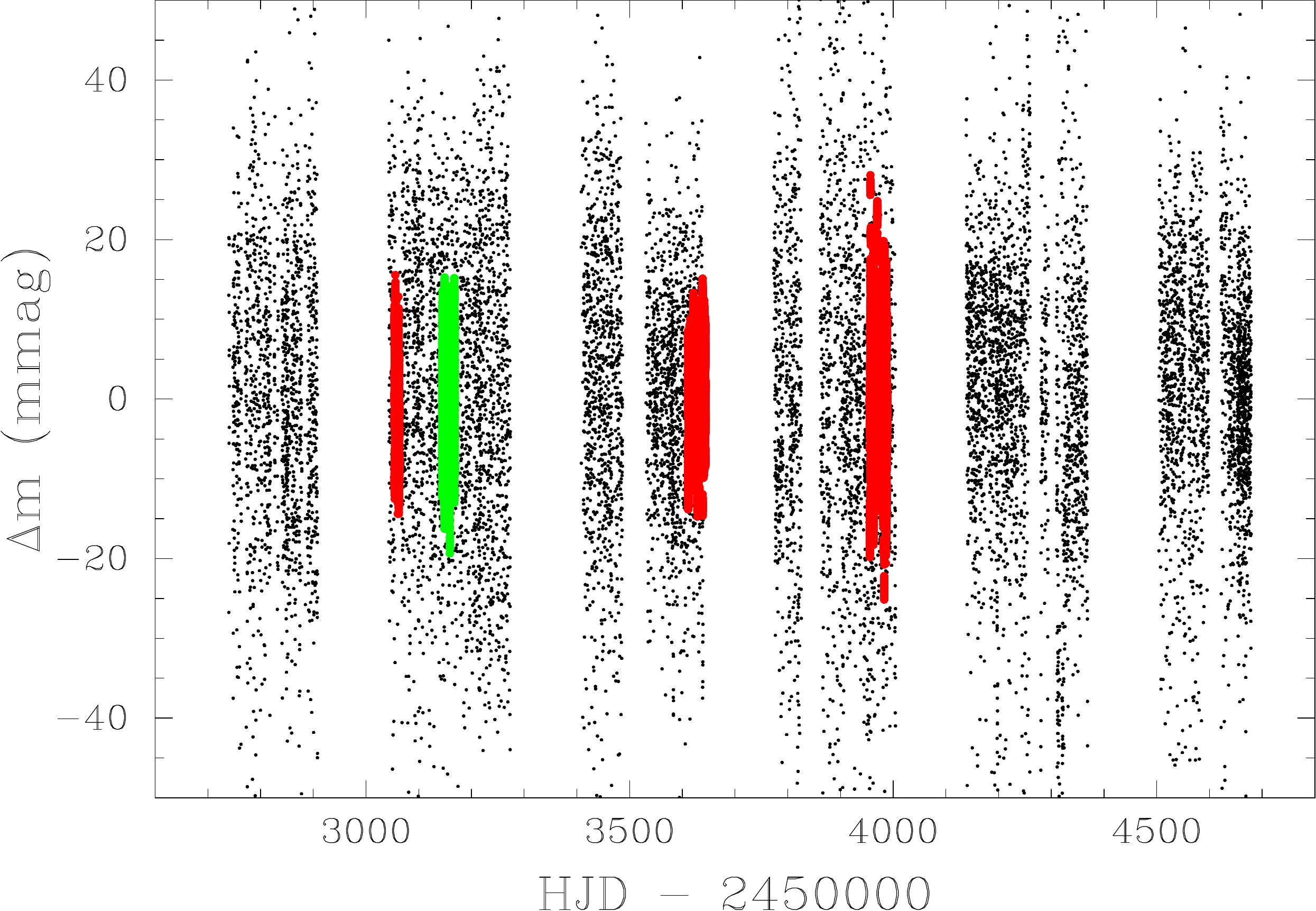}
\caption{Time distribution of $\zeta$~Oph data.  The extensive dataset
  is the \emph{SMEI} results; the four shorter sequences, showing 
  smaller dispersions in magnitude, are 
results from
\emph{WIRE} (red) and
  \emph{MOST} (green, second cluster of points).}
\label{fig:timeseries}
\end{figure}

\begin{figure}
\centering
\includegraphics[scale=0.45]{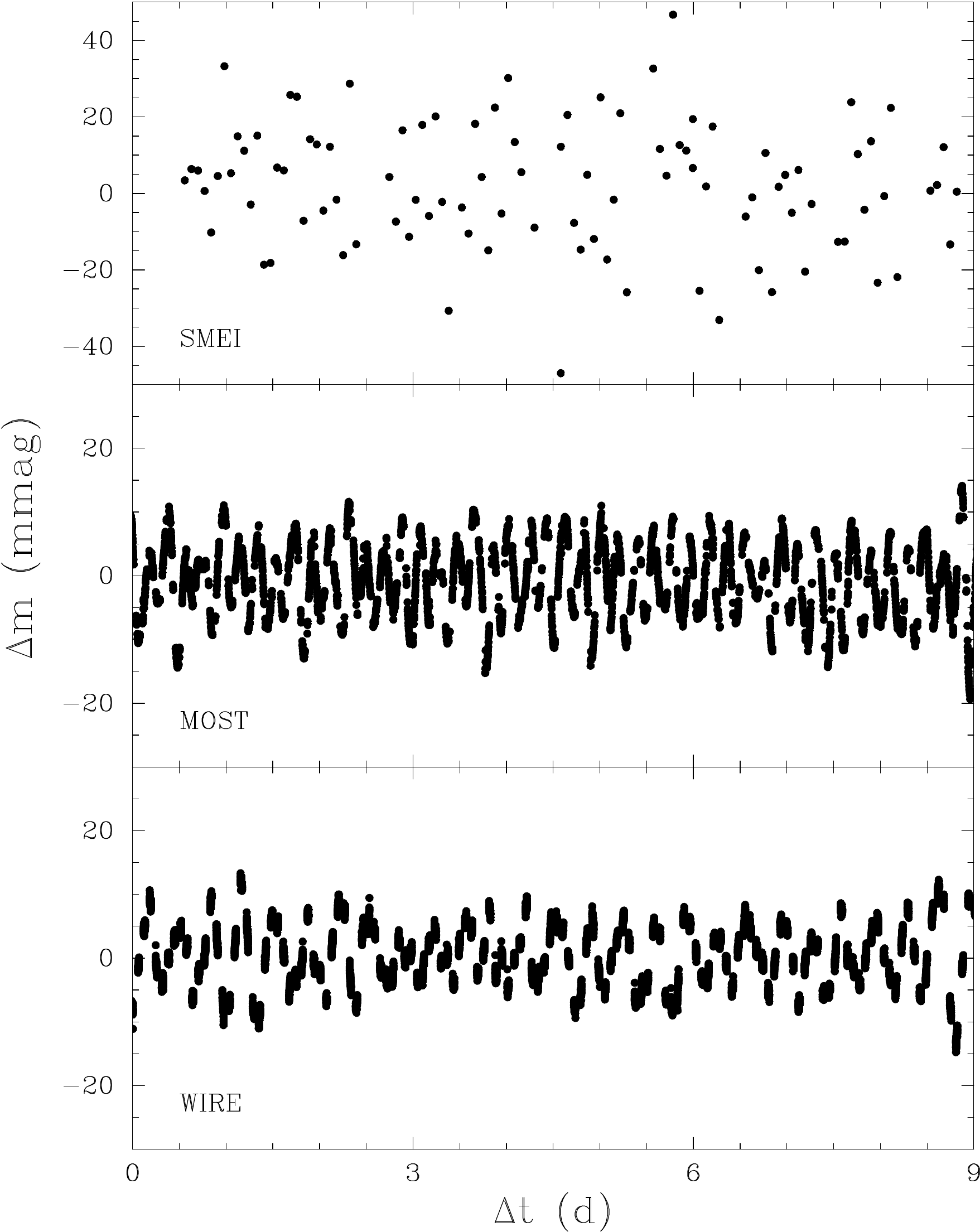}
\caption{Illustrative data sequences, starting
HJD~2\,453\,150 (\emph{SMEI, MOST}) and
2\,453\,620 (\emph{WIRE}).}
\label{fig:timeseries_segments}
\end{figure}

An overview of the time sampling is provided by Table~\ref{tab:dates}
and Fig.~\ref{fig:timeseries}, while the data quality is illustrated
in Fig.~\ref{fig:timeseries_segments}.  Although of somewhat lower
cadence and accuracy than the other datasets, the \emph{SMEI}
observations are noteworthy in that they span six years, with
approximately eight months' almost continuous coverage annually, allowing us
to examine the long-term behaviour of periodic signals, presumed to
arise from pulsations.

\begin{table*}
\centering
\caption{Summary of observations.}
\begin{tabular}{c l r | c l r}
\hline
Satellite & 
\multicolumn{1}{c}{Observation Period} & 
\multicolumn{1}{c}{$N$} & Satellite & 
\multicolumn{1}{c}{Observation Period} & \multicolumn{1}{c}{$N$} \\
\hline
\emph{SMEI} & 2003 Feb 10 -- Sept 25 & 1467 & \emph{WIRE} & 2004 Feb 18 -- Feb 27 & 6663   \\
\emph{SMEI} & 2004 Feb 7 -- Sept 25 & 2473  & \emph{WIRE} & 2005 Aug 28 -- Sept 30 & 40739 \\
\emph{SMEI} & 2005 Feb 6 -- Sept 25 & 2255  & \emph{WIRE} & 2006 Aug 8 -- Sept 9 & 28113   \\
\emph{SMEI} & 2006 Feb 6 -- Sept 26 & 2194  \\
\emph{SMEI} & 2007 Feb 7 -- Sept 25 & 2340  & \emph{MOST} & 2004 May 18 -- June 11 & 9084   \\
\emph{SMEI} & 2008 Feb 7 -- Aug 1 &  2101    \\
\hline                                        
\end{tabular}
\label{tab:dates}
\end{table*}

\subsection{\emph{SMEI}}
\label{sec:smei}

\emph{SMEI} was one of two instruments on the \emph{Coriolis}
satellite, launched on 2003 January 6; data acquisition ceased on 2011
Sept 28.  Designed to detect and forecast coronal mass ejections
moving towards the Earth, \emph{SMEI} had three imaging cameras, but
camera~3 suffered a relatively high-temperature state, and as a
result the quality of its photo\-metric data is relatively poor.  Here
we only use results from cameras~1 and~2.

\emph{SMEI} was capable of measuring millimagnitude variability
down to $\sim6{\fm}5$.  
The optical system was unfiltered, so the passband was dominated by
the spectral response of the CCDs: the quantum efficiency peaked at
45\%\ at 700~nm, falling to 10\%\ at $\sim$460 and 990~nm.  The
cameras each had a field of view of 60$^{\circ}$ $\times$ 3$^{\circ}$,
and were mounted such that they scanned nearly the entire sky every
101 minutes.  The duty cycle for the $\zeta$~Oph time series is
46.6\%, a typical value for \emph{SMEI} photo\-metry.  Our $\zeta$~Oph
analysis uses data from six seasons, spanning $\sim5\nicefrac{1}{2}$~years
(Table~\ref{tab:dates}), after which there is a falloff in data quality.

The \emph{SMEI} instrument is fully described by \citet{Eyles03},
and a brief description of the data reduction
can be found in \citet{Spreckley08};
other \emph{SMEI}-based photo\-metric
investigations include studies of $\alpha$~Boo,
$\beta$~UMi,
$\gamma$~Dor,
$\alpha$~Eri, Cepheid variables, and
the magnetic CP star CU~Vir 
\citep{Tarrant07, Tarrant08a, Tarrant08b,
Goss11,
Berdnikov10, Pyper13}.

\subsection{\emph{WIRE}}
\label{sec:wire}

The \emph{WIRE} satellite was launched
in 1999.  Its main infra\-red camera never came into operation due to
loss of coolant soon after launch, but the star tracker was
successfully employed from 1999 to 2006 to measure precise
light-curves of bright stars, in a passband roughly corresponding to
$V+R$, determined by the CCD response
\citep{Bruntt06, Bruntt08}. \emph{WIRE} observed
$\zeta$~Oph in three runs, and we extracted photo\-metry using the
\emph{WIRE} pipeline \citep{Bruntt05}.

\subsection{\emph{MOST}}
\label{sec:most}

The \emph{MOST} satellite is a photo\-metric instrument dedicated to
asteroseismic observations \citep{Walker03}, and again had a broad
spectral response ($\sim$350--700nm). \emph{MOST} observed $\zeta$~Oph
for 23 days in 2004;  these observations have already
been discussed in detail by \citet{Walker05}.

\section{Time-series Analysis}
\label{sec:data_analysis}

\begin{figure}
\centering
\includegraphics[scale=0.52,angle=270]{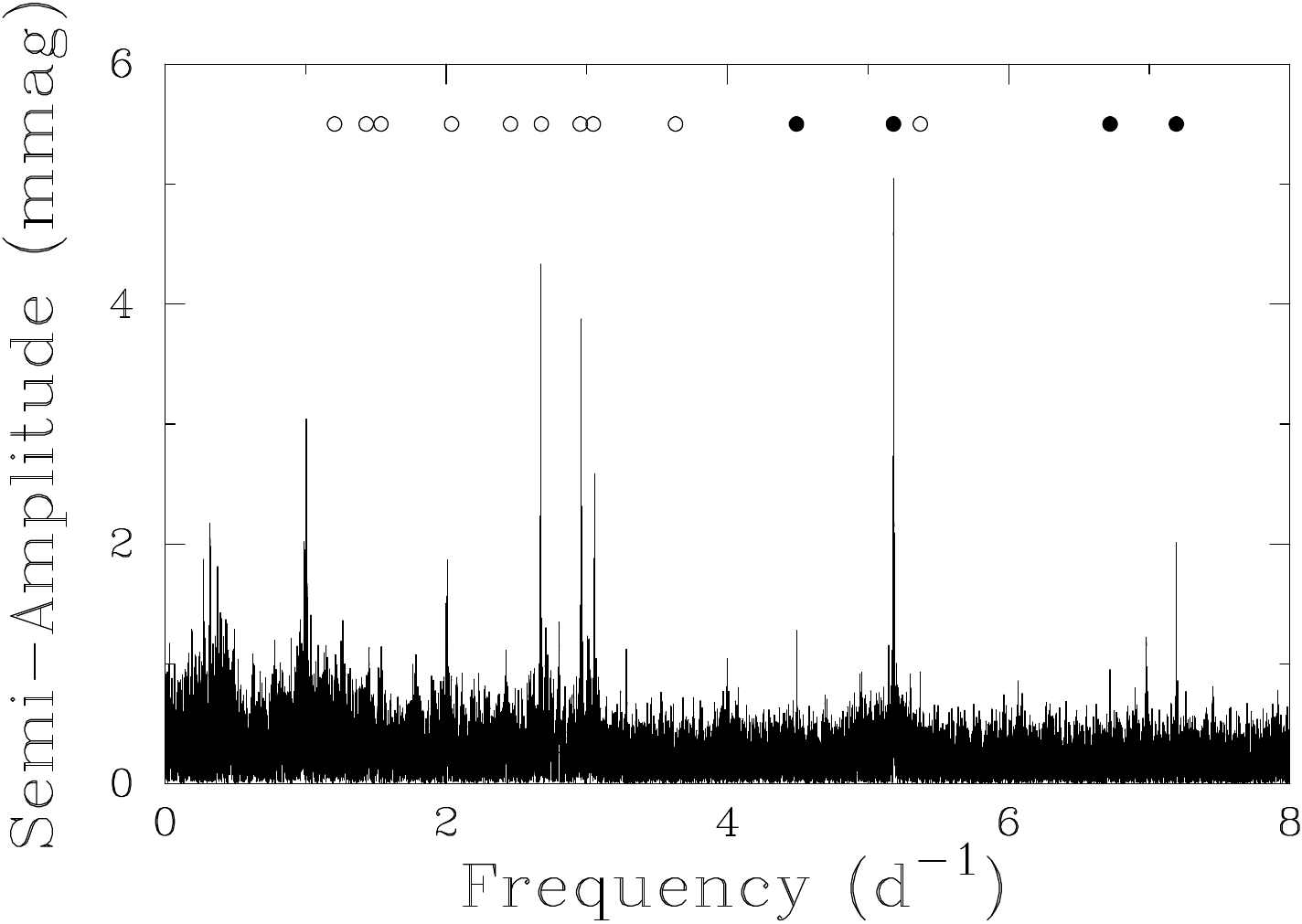}
\caption{Amplitude spectrum of $\zeta$~Oph from \emph{SMEI}
  photometry.  The open circles indicate frequencies reported by
  \citet{Walker05} from \emph{MOST} photometry;  filled circles also
  have spectro\-scopic identifications \citep{Reid93, Kambe97,
    Walker05}.  Signals at integer multiples of 1~\pd\ are assumed not
  to be
  astrophysical in origin.}
\label{fig:smei_spec}
\end{figure}

All \emph{SMEI} photo\-metry shows long-term variations of
instrumental origin (e.g., \citealt{Goss11}).  These were removed with
a ten-day running-mean filter, and a time-series analysis performed on
the corrected data using {\sc Period04} \citep{Lenz05}.
Fig.~\ref{fig:smei_spec} shows the date-corrected
discrete-fourier-transform amplitude spectrum (\citealt{Ferraz81});
although the formal Nyquist frequency imposed by the orbital period is
7.086~\pd, the window function is very clean, and useful information
can be extracted at somewhat shorter periods (into the frequency domain
explored by spectro\-scopic investigations).  However, it is clear from
Fig.~\ref{fig:smei_spec} (and from other \emph{SMEI}-based analyses)
that the Sun-synchronous orbit of the satellite generates signals at
frequencies of 1~d$^{-1}$ and multiples thereof.  Any astrophysical
signals which occur at these frequencies cannot be reliably
identified in the \emph{SMEI} data alone.

\subsection{Summary of frequencies}
\label{sec:frq}

`Significant' astrophysical signals (those with S/N$\ge$4 in the full
\emph{SMEI} dataset) are summarised in Table~\ref{tab:freq}, where the
tabulated errors on the frequencies and amplitudes have been
calculated from Monte-Carlo simulations.  The 7.19~\pd\ signal
identified spectro\-scopically by \citet{Reid93} and \citet{Kambe97},
from observations obtained in 1989 and 1993, respectively, is present
in the photometry (and has therefore persisted for, or recurred over, 
two decades), but none of the longer-period signals they report
is recovered, with upper limits of $\sim$0.1--0.2~mmag.
\citet{Walker05} report additional signals at 4.49, 5.18, and
6.72~\pd\ in their spectro\-scopy (and \emph{MOST} photometry);
signals at these frequencies are also present in the \emph{SMEI}
results, although the last two are detected at only 2--3$\sigma$
significance.

The \emph{WIRE} and \emph{MOST} time series have been analysed in the
same way, with results included in Table~\ref{tab:freq}.  The
frequencies found in all three datasets are generally in good
agreement, though not all frequencies are detectable at all epochs.
In a few cases there are formally statistically significant
differences in frequencies from different datasets, but it is not
clear that these are astrophysically significant.  For example, the
5.18~\pd\ frequency in the 2006 \emph{WIRE} dataset appears to be
marginally lower than found in the full \emph{SMEI} and \emph{MOST}
results, but the 2006 \emph{SMEI} data alone, although of poorer quality
than the \emph{WIRE} results, support a higher value.  Our interpretation
of the data is, therefore, that there is no compelling evidence for
variations in \emph{frequency} for a given signal.  
To support this view, we
show the power spectra season by season in
Fig.~\ref{fig:split_spec}.   Essentially identical frequencies recur
each year, but with large variations in amplitude.

\subsection{Amplitude variability}
\label{sec:freq}


\begin{table*}
\centering
\caption{Signals identified with $\gtrsim$4-$\sigma$ confidence at frequencies
1--8~\pd.  (The Nyquist frequency for the \emph{SMEI} photometry 
is 7.1~\pd.)}
\begin{tabular}{l l|l l| l l | l l | l l |}
\hline
\multicolumn{2}{|c|}{\emph{SMEI} 2003--8}& \multicolumn{2}{|c|}{\emph{WIRE} 2004} & \multicolumn{2}{|c|}{\emph{MOST} 2004} & \multicolumn{2}{|c|}{\emph{WIRE} 2005} & \multicolumn{2}{|c|}{\emph{WIRE} 2006} \\
\multicolumn{1}{c}{Frequency} & \multicolumn{1}{c}{Amplitude} &\multicolumn{1}{c}{Frequency} & \multicolumn{1}{c}{Amplitude} & \multicolumn{1}{c}{Frequency} & \multicolumn{1}{c}{Amplitude} & \multicolumn{1}{c}{Frequency} & \multicolumn{1}{c}{Amplitude} & \multicolumn{1}{c}{Frequency} & \multicolumn{1}{c}{Amplitude} \\
\multicolumn{1}{c}{(d$^{-1}$)} & \multicolumn{1}{c}{(mmag)} &\multicolumn{1}{c}{(d$^{-1}$)} & \multicolumn{1}{c}{(mmag)} & \multicolumn{1}{c}{(d$^{-1}$)} & \multicolumn{1}{c}{(mmag)} & \multicolumn{1}{c}{(d$^{-1}$)} & \multicolumn{1}{c}{(mmag)} & \multicolumn{1}{c}{(d$^{-1}$)} & \multicolumn{1}{c}{(mmag)} \\
\hline
&&1.165(2) & 1.77(5) & \multicolumn{1}{c}{--} &     \multicolumn{1}{c}{--} &     \multicolumn{1}{c}{--} &     \multicolumn{1}{c}{--} &     \multicolumn{1}{c}{--} &     \multicolumn{1}{c}{--} \\
&&\multicolumn{1}{c}{--} &     \multicolumn{1}{c}{--} &     \multicolumn{1}{c}{--} &     \multicolumn{1}{c}{--} &     \multicolumn{1}{c}{--} &     \multicolumn{1}{c}{--}       & 1.2537(2) &   2.82(3) \\
&&\multicolumn{1}{c}{--}       & \multicolumn{1}{c}{--} &   \multicolumn{1}{c}{--} &     \multicolumn{1}{c}{--} &     \multicolumn{1}{c}{--} &     \multicolumn{1}{c}{--}       & 1.3549(3) &   2.02(3) \\
&&\multicolumn{1}{c}{--} &     \multicolumn{1}{c}{--} &     \multicolumn{1}{c}{--} &     \multicolumn{1}{c}{--} &     \multicolumn{1}{c}{--} &     \multicolumn{1}{c}{--}       & 1.4480(2) & 3.40(3) \\
&&\multicolumn{1}{c}{--} &     \multicolumn{1}{c}{--} &     \multicolumn{1}{c}{--} &     \multicolumn{1}{c}{--}       & 2.0385(2) & 1.13(1) &      \multicolumn{1}{c}{--} &     \multicolumn{1}{c}{--} \\
&&\multicolumn{1}{c}{--} &     \multicolumn{1}{c}{--} &     \multicolumn{1}{c}{--} &     \multicolumn{1}{c}{--} &     \multicolumn{1}{c}{--} &     \multicolumn{1}{c}{--} & 2.3816(4) & 1.54(3) \\
&&\multicolumn{1}{c}{--} &     \multicolumn{1}{c}{--} &     \multicolumn{1}{c}{--} &     \multicolumn{1}{c}{--} & 2.4334(3) & 0.90(1) &    2.4297(2) & 2.55(3) \\
&&\multicolumn{1}{c}{--} &     \multicolumn{1}{c}{--} &     \multicolumn{1}{c}{--} &     \multicolumn{1}{c}{--} &     \multicolumn{1}{c}{--} &     \multicolumn{1}{c}{--}       & 2.6269(2) & 2.43(3) \\
2.67137(2)&4.3(3)&2.648(1) & 2.16(5) & 2.6762(4) & 2.24(4) & 2.6706(1) & 1.86(1) & 2.6800(1) & 4.63(3) \\
&&\multicolumn{1}{c}{--} &     \multicolumn{1}{c}{--} &     \multicolumn{1}{c}{--} &     \multicolumn{1}{c}{--} &     \multicolumn{1}{c}{--} &     \multicolumn{1}{c}{--}       & 2.7014(1) & 4.31(3) \\
&&\multicolumn{1}{c}{--} &     \multicolumn{1}{c}{--} &
\multicolumn{1}{c}{--} &     \multicolumn{1}{c}{--} &
\multicolumn{1}{c}{--} &     \multicolumn{1}{c}{--}       & 2.8060(2)
& 3.51(3) \\
2.96041(2)&4.0(3)&\multicolumn{1}{c}{--} &     \multicolumn{1}{c}{--}    & 2.953(1) &    0.76(4) & 2.96120(5) & 5.32(1) & 2.9581(3) & 2.31(3) \\
&&\multicolumn{1}{c}{--} &     \multicolumn{1}{c}{--} & 3.0153(8) & 1.11(4) &    \multicolumn{1}{c}{--} &     \multicolumn{1}{c}{--} &     \multicolumn{1}{c}{--} &     \multicolumn{1}{c}{--} \\
3.05498(3)&5.1(3)&3.085(2) &      1.49(5) & 3.0481(6) & 1.50(4) &   3.0557(3) & 0.90(1) & 3.0578(4) & 1.51(3) \\
&&\multicolumn{1}{c}{--} &     \multicolumn{1}{c}{--} &     \multicolumn{1}{c}{--} &     \multicolumn{1}{c}{--}       & 3.7588(2) & 1.16(1) &      \multicolumn{1}{c}{--} &     \multicolumn{1}{c}{--} \\
4.49194(5)&1.3(2)&\multicolumn{1}{c}{--} &     \multicolumn{1}{c}{--}       & 4.4906 (8) & 1.07(4) &     \multicolumn{1}{c}{--} &     \multicolumn{1}{c}{--} &     \multicolumn{1}{c}{--} &     \multicolumn{1}{c}{--} \\
&&\multicolumn{1}{c}{--}       & \multicolumn{1}{c}{--} &   \multicolumn{1}{c}{--} &     \multicolumn{1}{c}{--} & 4.7092(5) & 0.53(1) &    \multicolumn{1}{c}{--} &     \multicolumn{1}{c}{--} \\
&&\multicolumn{1}{c}{--} &     \multicolumn{1}{c}{--} &     \multicolumn{1}{c}{--} &     \multicolumn{1}{c}{--} & 4.8650(5) & 0.53(1) &    \multicolumn{1}{c}{--} &     \multicolumn{1}{c}{--} \\
5.18082(1)&5.1(3)&5.1796(5) & 6.68(5) & 5.1805(1) & 7.22(4) &       \multicolumn{1}{c}{--} &     \multicolumn{1}{c}{--} &     5.1760(8) & 1.32(3) \\
$[$5.371(2)&0.9(4)$]$   &\multicolumn{1}{c}{--} &     \multicolumn{1}{c}{--} & 5.371(1) & 0.70(4) &     \multicolumn{1}{c}{--} &     \multicolumn{1}{c}{--} &     \multicolumn{1}{c}{--} &     \multicolumn{1}{c}{--} \\
$[$6.719(8)&1.0(3)$]$&\multicolumn{1}{c}{--} &     \multicolumn{1}{c}{--}       & 6.7209(7) & 1.28(4) &      \multicolumn{1}{c}{--} &     \multicolumn{1}{c}{--} &     \multicolumn{1}{c}{--} &     \multicolumn{1}{c}{--} \\
7.19196(5)&2.0(3)&7.205(4)&1.6(1)&7.196(3)&0.85(9)&7.1917(7)&0.83(3)&7.20(2)&0.9(1)\\
\hline
\end{tabular}
\label{tab:freq}
\end{table*}

\begin{figure*}
\centering
\includegraphics[scale=0.95]{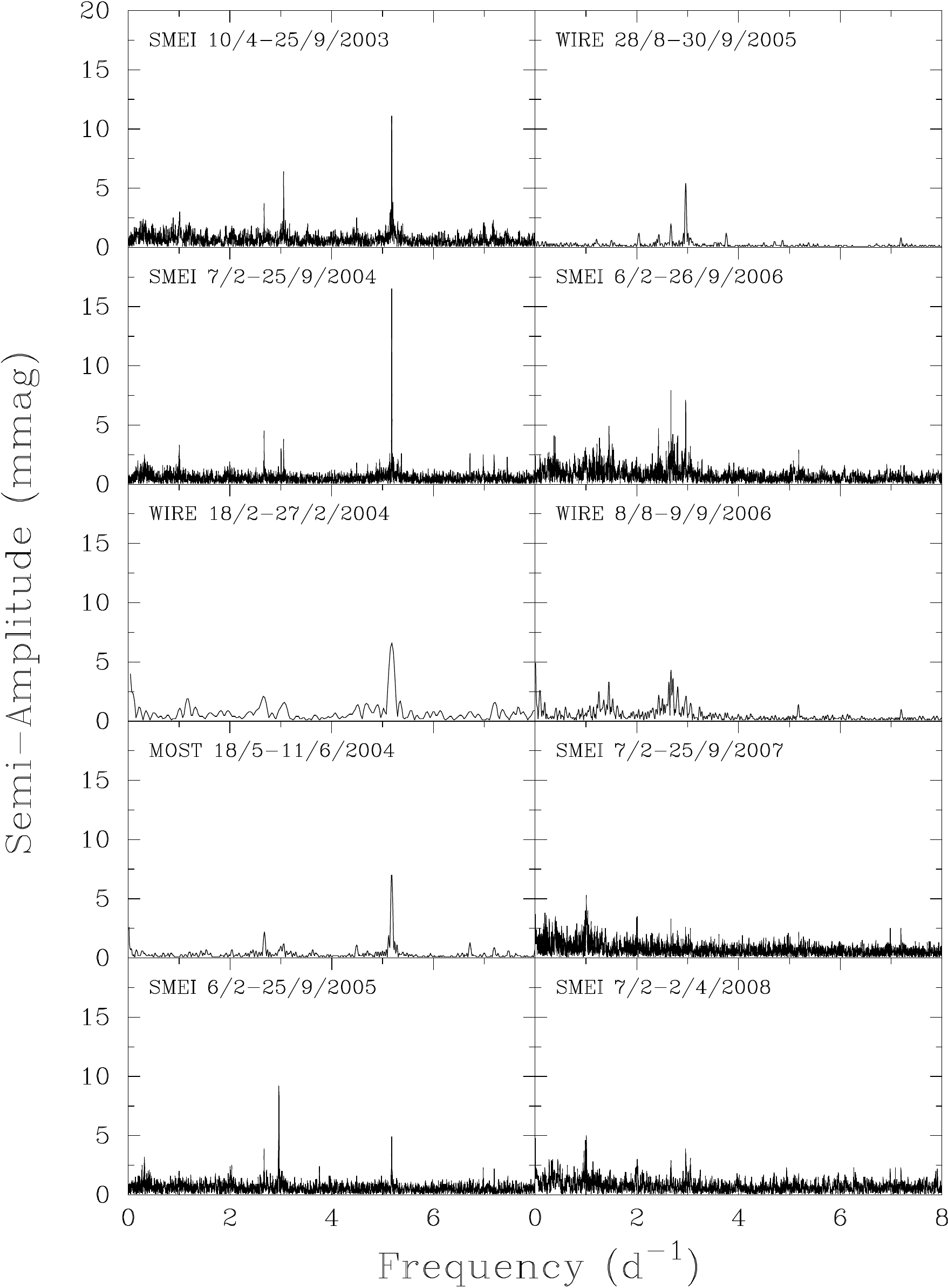}
\caption{Power spectra by observing season, ordered chronologically by
date of first observation.}
\label{fig:split_spec}
\end{figure*}

\begin{figure*}
\centering
\includegraphics[scale=0.80,angle=270]{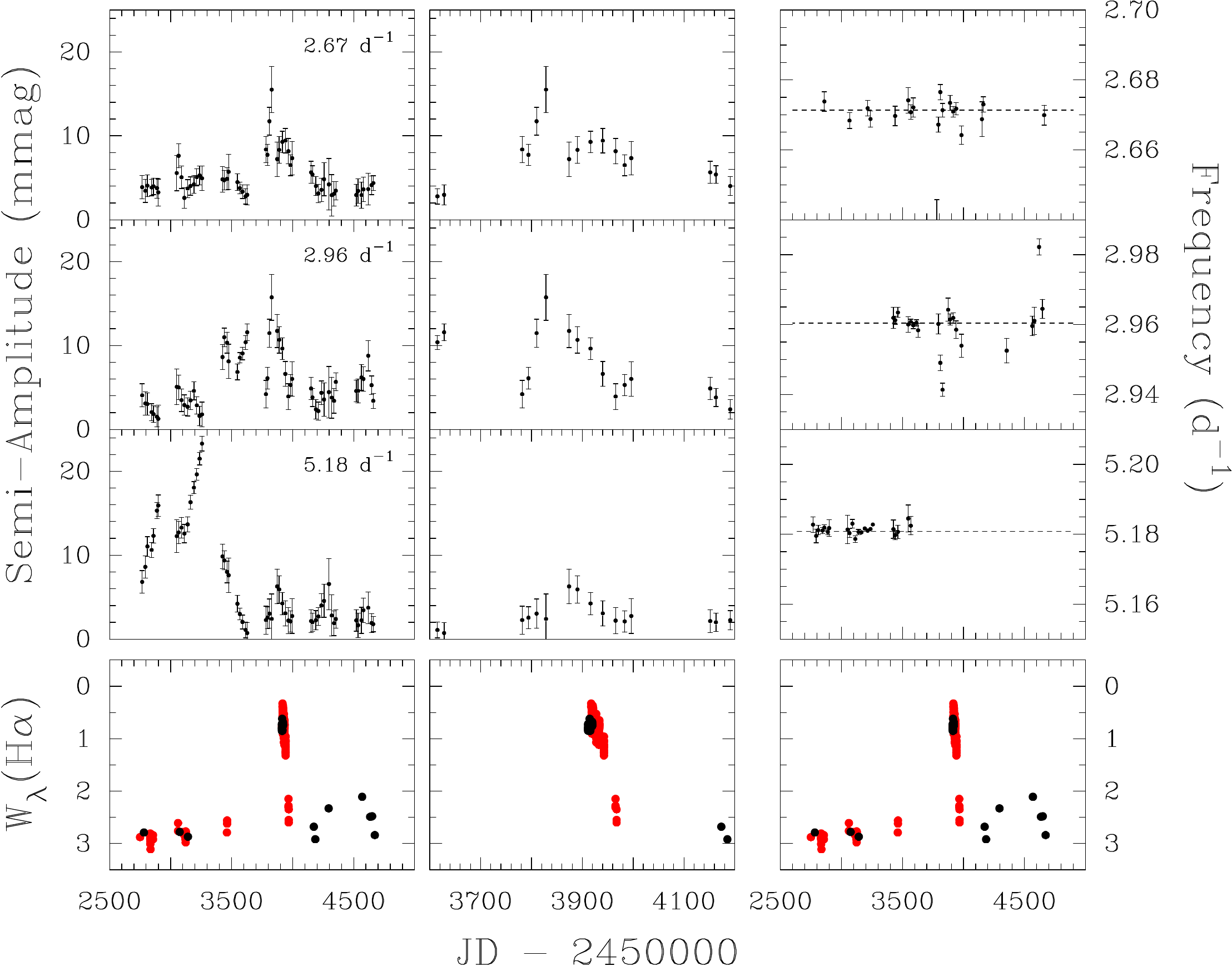}
\caption{(Left, top three panels) amplitudes for the strongest
  signals in the \emph{SMEI} dataset, evaluated for 50-d data
  segments at 25-d intervals;  (centre) zoom of data during the 2006
  emission-line episode.
(Right) frequencies for dates when the
  amplitude was $\ge{4}\times$ its error, where the errors
 have been estimated analytically, following
  \citeauthor{Montgomery99} (\citeyear{Montgomery99};  these are lower limits to true uncertainties).
Dashed horizontal lines indicate the mean frequencies determined from
the full \emph{SMEI} dataset.
The bottom panels show our H$\alpha$ equivalent-width measurements
(red: Ond\v{r}ejov spectra;  black: BeSS database); the cluster of points around
JD$\ldots$3930 with $W_\lambda \lesssim 2$\AA\ corresponds to an
emission-line episode.}
\label{fig:amp_phase_changes}
\end{figure*}

We take advantage of the long, uniform time series obtained by
\emph{SMEI} to examine these changes in amplitude in greater detail.
Fig.~\ref{fig:amp_phase_changes} shows the semi-amplitudes (and periods)
determined from \mbox{50-d} data segments, at 25-d
steps, for the strongest signals in the \emph{SMEI} photometry.  As
already evident from Fig.~\ref{fig:split_spec}, variability in signal
amplitude on timescales of order hundreds of days is the norm; most
clearly, the signal at 5.18~\pd\ has a large amplitude during 2003 and
2004, but becomes practically undetectable over the course of the 2005 observing
season, remaining at a very low level for the remainder of the period
under consideration.
(These conclusions are not artefacts of the data; the noise level and
fill factor of the \emph{SMEI} data show no important changes over
this time interval, and the \emph{WIRE} results exhibit the same trends,
though in less detail.)

In principle, it would be of obvious interest to investigate the phase
stability of the signals;  in practice, the phasing errors
across the extensive timespan studied here are too large to allow firm
conclusions in this regard, other than to state that there is no
evidence for significant phase drift.

\section{Discussion}
\label{sec:discussion}

\subsection{Signal duration}
The strongest signal found here, at 5.18~\pd\ (4{\fh}63), was first
reported by \citet{Balona92}, whose ground-based Str\"{o}mgren
photometry indicated a semi-amplitude of $\sim$0{\fm}01 in
1985.  He was unable to recover
this period in subsequent, relatively sparse, observations from the
1987--90 seasons, concluding both that there was no periodicity that
lasted more than one season, and that the short-period variations were
not stable.

The extensive, high-quality satellite photometry now available allows
us to revise these conclusions; signal amplitudes are indeed strongly
variable, on timescales $\sim{O}({10^2\mathrm{ d}})$, but, while
undetectable at some epochs, the same periods may be recoverable in
datasets separated by two decades.  \citet{Balona92} also found a
2.66~\pd\ signal (among other tentative identifications) in 1989
observations; this very probably matches the 2.67~\pd\ signal observed
in the \emph{SMEI}, \emph{WIRE} and \emph{MOST} datasets, again
emphasizing that at least some signals may be present over decades
(cf.\ 7.19~\pd; $\S$\ref{sec:frq}), though whether they persist
continuously, sometimes below detection thresholds, remains moot.

\subsection{Emission-line episode}

Spectro\-scopy that is contemporaneous with our photometry is available
in the BeSS archive (\citealt{Neiner11}; dispersions of
$\sim$0.1--0.3~\AA/pixel) and from Ond\v{r}ejov Observatory (Harmanec,
personal communication; $\sim$0.25~\AA/pixel);  the formal
signal-to-noise ratios are typically a few hundred per sample.  We have corrected
these spectra for absorption in the Earth's atmosphere by division by a scaled
high-resolution reference telluric spectrum before measuring
equivalent widths.

The data show that $\zeta$~Oph under\-went an emission episode in
summer 2006, similar to that illustrated by \citeauthor{Ebbets81}
(\citeyear{Ebbets81}, his Fig.~1).  Our H$\alpha$ equivalent-width
measurements are included in Fig.~\ref{fig:amp_phase_changes}.  There
is a suggestion that the emission-line episode may have been
accompanied (or slightly preceded) by simultaneous small increases in
the amplitudes of the 2.67, 2.96, and 5.18~\pd\ signals.
Unfortunately, with only one known emission-line episode during the course
of our observations, it is not possible to draw firm conclusions from
this coincidence; nevertheless, it is suggestive that the two lower
frequencies attained the greatest amplitudes recorded in our
photometry at that time.

The largest amplitudes, and largest changes in amplitude, are recorded for the
higher-frequency 5.18~\pd\ photometric signal, and there is no clear
association between that signal and any emission-line activity.
However, \citet{Walker05} showed that this frequency could very
plausibly correspond to a first-overtone \emph{radial} mode
(exciting non-radial modes that give rise to the
spectro\-scopic line-profile variability).  The potential pulsation
mechanism for producing decretion disks discussed by
\citeauthor{Cranmer09} (\citeyear{Cranmer09}; see also
\citealt{Ando86}) relies on the injection of angular momentum into
the upper atmosphere by \emph{non-radial} modes.   
Although there are as yet no mode identifications for the 2.67 and
2.96~\pd\ signals (which do not have published spectroscopic counterparts),
it therefore remains plausible that the emission-line episode could be causally
associated with increases in amplitudes of non-radial modes.

\section{Conclusion}
\label{sec:conclusion}

Data from the \emph{SMEI, WIRE, \emph{and} MOST} satellites have been
analysed to investigate periodic signals in broad-band optical
photometry of the Oe star $\zeta$~Oph obtained over a span of almost
6~years.  We confirm multiperiodic variability at the $\sim$10~mmag
level and, for the first time, track systematic changes in signal
amplitudes on timescales of order $\sim{10}^2$~d; some signals, while
not continuously detectable, are never\-the\-less present in
observations separated by 20~years.  
There is tentative evidence of a photometric
signature of the 2006 emission-line episode;
although no direct
correspondence is evident between overall photometric and
spectro\-scopic activity, this \emph{may} reflect different roles of
radial and non-radial modes in the formation of a decretion disk.

\section*{Acknowledgements}
\label{sec:acknowledgements}

KJFG, IRS, WJC, and YE acknowledge the support of STFC.  IDH is a
Jolligoode Fellow.  This work has made use of the BeSS database,
operated at the Observatoire de Meudon; we thank the contributing
observers (C.~Buil, J.~Guarroflo, C.~Neiner, E.~Pollmann, and O.~Thizy),
together with Petr Harmanec, who kindly provided spectra from
Ond\v{r}ejov, and Steve Spreckley, who assisted with the \emph{SMEI}
data reduction.  Our anonymous referee's remarks were constructively
stimulating.

\bibliographystyle{mn2e}

\bibliography{zeta}

\appendix

\label{lastpage}

\end{document}